\documentclass[12pt]{article} 

\usepackage{mathtools, amsfonts, amsthm, latexsym, amssymb,dsfont}
\usepackage[T1]{fontenc}
\usepackage[tracking=true]{microtype}
\usepackage{lmodern}
\usepackage{verbatim}

\usepackage[margin=1in]{geometry}

\usepackage{verbatim}
\usepackage{graphicx}
\usepackage{subcaption}
\usepackage{booktabs}
\usepackage{longtable}
\usepackage{cite}

\usepackage{color}
\definecolor{darkblue}{cmyk}{0.9,0.9,0,0}
\definecolor{wine-stain}{rgb}{0.5,0,0}
\usepackage[colorlinks, allcolors=darkblue, linktoc=all]{hyperref} 
\usepackage{setspace}

\renewcommand{\L}{\mathcal L}

\newcommand{\beq}{\begin{equation}}
\newcommand{\eeq}{\end{equation}}
\newcommand{\be}{\begin{equation}}
\newcommand{\ee}{\end{equation}}
\newcommand{\bea}{\begin{align}}
\newcommand{\eea}{\end{align}}

\def\gsim{ \lower .75ex \hbox{$\sim$} \llap{\raise .27ex \hbox{$>$}} }
\def\lsim{ \lower .75ex \hbox{$\sim$} \llap{\raise .27ex \hbox{$<$}} }

\def\beq{\begin{eqnarray}}
\def\eeq{\end{eqnarray}}

\def\p{{\cal P}}

\def\L*{{\cal L}_*}
\def\L{\mathcal{L}}
\def\({\left(}
\def\){\right)}

\def\nn{\nonumber}
\def\p{\partial}

\def\p{\partial}

\def\<{\langle}
\def\>{\rangle}

\def\xyma{\xymatrix@M.7em}
\def\xymas{\xymatrix@M.1em}
\newcommand{\ba}{\begin{eqnarray}}
\newcommand{\ea}{\end{eqnarray}}

\begin{document}
\thispagestyle{empty}
\vspace*{0.5in} 
\begin{center}
{\Huge Continuity, Localization, and Cosmology}
\vskip 0.25cm
{\Huge in Warped Geometry}

\vskip 0.7cm
\centerline{
{\large {Gregory Gabadadze,}}
{\large {Daniel Older}}
{\large {and David Pirtskhalava}}
}
\vspace{.2in} 

\centerline{{\it Center for Cosmology and Particle Physics, Department of Physics}}
\centerline{{\it New York University, New York, NY, 10003, USA}}
\end{center}

\begin{abstract}

This is the first of two papers studying localization of massive bulk fields on a bane in 5D anti-de Sitter spacetime, 
and some of their cosmological consequences. Here we focus on a massive 5D scalar, which is known to lack 
a localized mode, and  discuss how a seeming discontinuity between  this theory and the massless theory -- 
known  to support a localized zero mode -- is resolved thanks to peculiar analytic  properties 
of  the massive two-point amplitude. Furthermore, 
we propose a boundary term that leads to the emergence of a massless localized mode in the massive theory.
Last but not least, we consider the case when the brane world-volume is de Sitter spacetime, and  prove the existence
of a localized massive mode. We discuss  how these results, taken collectively, can be used to describe the 
accelerated expansion due to the massive 5D scalar field in an early,  or in a late-time universe.

\end{abstract}

\newpage
\noindent

\section{Introduction and Summary}

Randall and Sundrum (RS) \cite {Randall:1999ee} have shown that a 5D massless 
graviton yields a localized massless 4D graviton in 5D AdS 
spacetime with a $Z_2$ symmetric brane; the rest of the Kaluza-Klein(KK) modes form a 
gapless continuum, with the wavefunctions for the lighter KK modes suppressed on the brane, 
leading to approximate recovery of  4D physics at low energies.\footnote{Note that $AdS_5$ refers to the 
universal cover of  5D AdS space-time \cite {Avis:1977yn}. There exists a stable 
solitonic domain wall  solution with 4D Poincare invariant 
worldvolume, and  $Z_2$ symmetry in the direction transverse to the worldvolume 
\cite{Behrndt:1999kz}.  This solution, in a certain approximation, can be thought of 
as infinitely thin brane with all its fluctuations becoming negligible for a low energy 
brane observer.}  The RS discovery has a  numerous physical and theoretical 
consequences, which are well known.

It turns out that massless 5D scalar also yields a localized massless 4D mode 
on the brane worldvolume \cite {Bajc:1999mh}, with the gapless  KK continuum 
and the KK wavefunctions identical to those of massless 5D graviton.
However, massless vectors and spinors  do not produce  localized massless 
modes in the minimal setup. 

Our goal is to consider whether massive spin-0, spin-1, and spin-2 fields can be localized
or quasilocalized  on a brane in $AdS_5$.  This is the first of the two papers on this topic; here 
we focus on a 5D massive scalar field to emphasize some of the subtleties which are 
independent of the spin, but also to delineate cosmological scenarios 
specific to such a massive 5D scalar.

The question of localization of a 5D massive field 
was first studied in \cite {Dubovsky:2000am}, where it was found that there is no 
localized mode irrespective how small the 5D mass is. Yet, there is a 
resonant mode, that can mimic physics of the zero mode in a certain approximation 
\cite{Dubovsky:2000am}. 

Seemingly, there is discontinuity between the massless and massive theory --
the  former  has a massless 4D mode in its spectrum while 
the latter does not,  irrespective how small the 5D scalar mass is. 
Indeed, this discontinuity is real as long as the  spectra of the modes are concerned --
the bulk mass  makes the wold-be localized mode non-normalizable, for any nonzero 
value of the mass. However, we  will show that there is no discontinuity 
in the two-point amplitude of the theory. This involves careful consideration 
of the analytic  properties of this amplitude,  as it is done in Section 2.

In Section 3 we  show how one can modify the brane worldvolume theory 
by adding a simple 4D term for the scalar to reinstate the massless 
4D localized mode for a massive 5D bulk scalar  field. The new term is a "tachyonic"
4D worldvolume  mass  for the scalar, which depends on the bulk mass. 
We should note, however,  that the 4D term is overwhelmed by the positive 5D 
mass term and nowhere one encounters any tachyonic instabilities in the theory. 
The new term  create an additional "attraction"  in a potential for the spectral problem 
for the KK modes,  and this is enough to reinstate the massless mode.

In Section 4 we consider 5D massive scalar in the background geometry with 4D de Sitter 
worldvolume.  We prove the existence of a localized massive mode in this case and a KK continuum
starting above the gap determined by the curvature of the 4D de Sitter spacetime. We then outline 
how such a geometry can be an approximation to the one on which 4D curvature is provided
by a slowly rolling 5D massive scalar field. We point out differences of such a scheme 
from the one  with 4D massive scalar slowly rolling and providing inflation or 
late time acceleration (quintessence).

\section{Massive Scalar in 5D}
\label{massivescalar}

\subsection*{The setup}
Following \cite {Dubovsky:2000am}, we consider a massive scalar field in the 5-dimensional anti de Sitter spacetime
endowed with ${\Large \bf Z_2}$ symmetry across its boundary, described by the following interval
\be
ds^2 = \Omega^2(z)\, \eta_{MN}\,  dx^M dx^N \,, \quad \Omega(z) = \frac{L}{L + |z|}\,,
\ee
where the mostly minus metric convention is assumed.~The coordinate $z$, parametrizing the fifth dimension ranges from $-\infty$ to $+\infty$ and the theory is constrained to be ${\Large \bf Z_2}$ invariant under the flip of sign $z\to -z$, under which the scalar is assumed to be even, $\phi(x, z) = \phi(x, -z)$\,.\,We are interested in the 4D effective theory on the physical brane, located at $z=0$, where $\phi$ is additionally coupled to a 4D source $j(x)$\,.~The complete action thus reads
\begin{align} \label{action}
S_\phi &=  \int d^4 x \int_{-\infty}^{\infty} d z \, \bigg[ \sqrt{-g}\, \Big(\,\frac{1}{2} g^{M N} \partial_{M} \phi \partial_N \phi - \frac{1}{2} m_5^2 \phi^2 \Big) +  L \phi(x) j(x)\delta(z)\bigg ] \\\nn
&= \int d^4 x \int_{-\infty}^{\infty} d z\, \Omega^3 \Big(\frac{1}{2} \partial_{\mu} \phi \,\partial_{\mu} \phi - \frac{1}{2} (\partial_z \phi )^2- \frac{1}{2} \Omega^2 m_5^2 \phi^2 \Big) +  \int d^4 x L \phi(x) j(x).
\end{align}
All 4D indices on the last line are assumed to be contracted with the flat 4D metric, and we do not distinguish between upper and lower 4D components. Varying this action with respect to $\phi$ yields the following equation of motion
\begin{align} \label{equation}
\Big(-\Box + \partial_z^2 - \frac{3 \text{sgn}(z)}{\vert z \vert + L} \partial_z - \frac{(m_5 L)^2}{(\vert z \vert + L)^2} \Big) \phi (x,z) = -L j(x) \delta(z)\,,
\end{align}
which, when integrated across the brane (that is, within the interval $z \in [-\epsilon, \epsilon]$, with $\epsilon \rightarrow 0$) implies the following boundary condition for $\phi$:
\begin{align} \label{boundarycondition}
\partial_z \phi \vert_{z = 0} = - \frac{L}{2}\, j(x)\,.
\end{align}
In order to derive the 4D effective action on the brane, we will need to solve the system \eqref{equation} and \eqref{boundarycondition}. This is the subject of the next subsection. 



\subsection*{Kaluza-Klein modes}

Consistently with the equation of motion \eqref{equation} and boundary condition \eqref{boundarycondition}, the 5D field $\phi$ can be decomposed in terms of the 4D KK modes as follows:
\begin{align} \label{KKdecomp}
\phi(x, z) = \int_0^{\infty} dm L \, \phi^{(m)}(x)\, \chi^{(m)}_{\nu} (z)\,,
\end{align}
where we have defined $$\nu = \sqrt{4+(m_5 L)^2}\,.$$~The KK wavefunctions $\chi^{(m)}_{\nu}$ satisfy the following bulk equation
\begin{align} \label{KKequation}
\Big(\partial_z^2 - \frac{3 \text{sgn}(z)}{\vert z \vert + L} \partial_z - \frac{(m_5 L)^2}{(\vert z \vert + L)^2} \Big) \chi^{(m)}_{\nu} (z) = -m^2\, \chi^{(m)}_{\nu} (z)\,,
\end{align}
complemented with the boundary condition at the brane
\begin{align} \label{KKboundarycondition}
\partial_z \chi^{(m)}_{\nu} (z) \vert_{z = 0} = 0\,.
\end{align}
Explicitly, these KK wavefunctions read \cite{Dubovsky:2000am}
\begin{align}
\chi^{(m)}_{\nu} (z) = \sqrt{\frac{m L}{2}} \(\frac{\vert z\vert + L}{L}\)^2\Big[a_m J_{\nu} \big(m \(\vert z \vert+L\)\big) + b_m Y_{\nu} \big(m \(\vert z \vert+L\)\big)\Big]\,,
\end{align}
where the two coefficients $a_m$ and $b_m$ are given by the following expressions
\begin{align}
a_m &= - \frac{Y_{\nu-1}(mL) - \frac{\nu - 2}{m L} Y_{\nu}(mL)}{\sqrt{\(Y_{\nu-1}(mL) - \frac{\nu - 2}{m L} Y_{\nu}(mL)\)^2 + \(J_{\nu-1}(mL) - \frac{\nu - 2}{m L} J_{\nu}(mL)\)^2}} \\\nn
b_m &= \frac{J_{\nu-1}(mL) - \frac{\nu - 2}{m L} J_{\nu}(mL)}{\sqrt{\(Y_{\nu-1}(mL) - \frac{\nu - 2}{m L} Y_{\nu}(mL)\)^2 + \(J_{\nu-1}(mL) - \frac{\nu - 2}{m L} J_{\nu}(mL)\)^2}} \,.
\end{align}
Importantly, for non-zero bulk mass $m_5$ there is no normalizable zero mode, that is, no mode with $m=0$
\cite {Dubovsky:2000am}.~(For $m_5$ strictly zero, such a mode does exist and 
would correspond to the $z$-independent bulk profile $\chi^{(0)}= \text{const}$ \cite{Bajc:1999mh}.)~Nevertheless, even for a non-zero $m_5$, the spectrum features a \emph{gapless} continuum of massive KK modes, whose wavefunctions satisfy the usual orthonormality conditions
\begin{align} \label{normalization}
\int_{-\infty}^{\infty} dz\, \(\frac{L}{\vert z\vert + L}\)^3 \chi^{(m)}_{\nu} (z) \chi^{(m')}_{\nu}(z) = \delta(m- m')\qquad (m, m' > 0)\,.
\end{align}
Plugging decomposition \eqref{KKdecomp} into the original 5D action \eqref{action}, one arrives at the 4D effective action for the KK continuum
\begin{align} \label{KKaction}
S_{\phi} = L &\int d^4 x  \int_0^{\infty} dmL\, \Big[ \frac{1}{2} (\partial \phi^{(m)})^2 - \frac{m^2}{2} (\phi^{(m)})^2 +  \chi^{(m)}(0)\, \phi^{(m)} j\Big]\,.
\end{align}
At each KK level, the canonically normalized 4D field $\phi^{(m)}/\sqrt{L}$ couples to the brane source $j$ with strength $\chi^{(m)}(0)/\sqrt{L}$. This coupling weakens for higher KK modes.

The force due to the exchange of massive KK modes between two four-dimensional brane sources is determined by the following amplitude (transformed to momentum space along the brane dimensions)
\begin{align}
\Delta_{\nu} (p^2) = \int_0^{\infty} dm L\, \frac{\vert \chi^{(m)}_{\nu} (0) \vert^2}{p^2 - m^2 + i\epsilon} = \frac{L^2}{2}\, \bigg[(p L)  \frac{H^{(1)}_{\nu-1} (p L)}{H^{(1)}_{\nu} (p L)} - (\nu - 2)\bigg]^{-1} \quad (\nu > 2)  \,,
 \label{propagator}
\end{align}
where $p\equiv \sqrt{p^2}$.~This formula, \emph{valid strictly for} $\nu > 2$, was first derived in Ref. \cite{Dubovsky:2000am}.~Addressing the case of $\nu = 2$, on the other hand, is subtle and will be our focus in
the remainder of this section.

\subsection*{Continuity in the bulk mass}
The main task of the present subsection is demonstrating continuity of the physical predictions of the theory in the $m_5\to 0$ limit.~This limit is non-trivial: as we have remarked above, depending on whether the parameter $m_5$ is \textit{strictly} zero or not, there is an extra zero mode present in the KK spectrum, leading to a discontinuity in the degrees of freedom in the 4D theory on the brane.

Ref.~\cite{Dubovsky:2000am} has demonstrated how exactly continuity occurs for non-zero, but small $m_5$ at the level of the Newtonian potential, experienced by brane sources due to the exchange of the scalar's KK modes.~The crucial role is played by a special resonant mode, composed of the KK modes of the 4D theory.~This resonance, which only exists for $m_5\neq 0$ (or $\nu\neq 2$),  
is characterized by mass and width, given by 
\begin{align}
m_0^2 \approx \frac{m_5^2}{2}\,, \qquad
\Gamma &\approx \frac{\pi}{16 \sqrt{2} L} (m_5 L)^3\,. \label{massandwidth}
\end{align}
(Notice that for reasonable values of the bulk mass, $m_5 \lesssim 1/L$, the resonance's width is parametrically smaller than its mass.)~The presence of the pole corresponding to the resonance \eqref{massandwidth} can be readily established by studying the analytic structure of the amplitude \eqref{propagator}, see Ref.~\cite{Dubovsky:2000am} for details.~The negative imaginary part of the pole tends to zero as $m_5\to 0$, or equivalently as $\nu \to 2$ (of course, in this limit the real part vanishes as well and the pole asymptotically merges with the origin of the complex plane).~Therefore, in the limit $\nu\to 2$ one can conveniently write  Eq.~\eqref{propagator} as 
\be
\Delta_{\nu} (p^2)\vert_{\nu\to 2} =\text{P.V.} \( \frac{L^2}{2} \frac{H_2^{(1)} (p L)}{(p L) H^{(1)}_{1} (p L)}\)- i \pi  \delta(p^2)\,, \label{fullamp}
\ee
where $\text{P.V.}\(\dots\)$ in the first term denotes the principal value of the expression in the parentheses, 
while the second term essentially provides the $i\epsilon$-prescription for the pole at $p^2 = 0$, corresponding to the $m_5\to 0$ limit of the resonance.~To summarize, the analytic structure of the expression in \eqref{fullamp} features the pole, corresponding to the $m_5\to 0$ limit of the 4D resonance scalar (responsible for the $r^{-1}$ piece in the static potential), as well as a branch cut, corresponding to the entire gapless KK continuum (which leads to the $r^{-3}$ piece in the static potential).

We'd like to compare the above expression for the amplitude with its counterpart in the theory with strictly zero bulk scalar mass.~To that end, one can study the KK spectrum just like we did for the massive bulk theory.~As remarked above, in addition to the KK continuum there is now the true zero mode in the spectrum, and the amplitude becomes:
\begin{align} \label{MasslessSpectralGreensFunction}
\Delta_{\nu=2} (p^2) &= \frac{1}{p^2 + i \epsilon} + \int_0^{\infty} dm L\, \frac{\vert \chi^{(m)}_2 (0) \vert^2}{p^2 - m^2 + i \epsilon} \nonumber \\
&= \frac{1}{p^2 + i \epsilon} + L^2 \int_0^{\infty} d x\, \, \frac{1}{(p L)^2 - x^2 + i \epsilon} \,\,  \frac{2}{\pi^2 x (Y_1 (x)^2 + J_1 (x)^2)} \,,
\end{align}
where the first term on the right hand side of the first line comes from the $m=0$ zero mode exchange, while the second term --- from the exchange of the $m>0$ KK modes.~With the help of a few identities involving Bessel/Hankel functions, the second line of \eqref{MasslessSpectralGreensFunction} can be massaged into an integral over (almost) the entire real axis
\begin{align} \label{MasslessSpectralGreensFunction1}
\Delta_{\nu=2} (p^2) =\frac{1}{p^2 + i \epsilon} +  \frac{L^2}{2\pi i} \, \lim_{\delta\to 0^+}\( \int_{-\infty}^{-\delta} d x +  \int_{\delta}^{+\infty} d x\)\, \frac{1}{(p L)^2 - x^2 + i \epsilon}\, \frac{H_2^{(1)} (x)}{H_1^{(1)}(x)}\,.
\end{align}
\begin{figure}
  \includegraphics[width=\linewidth]{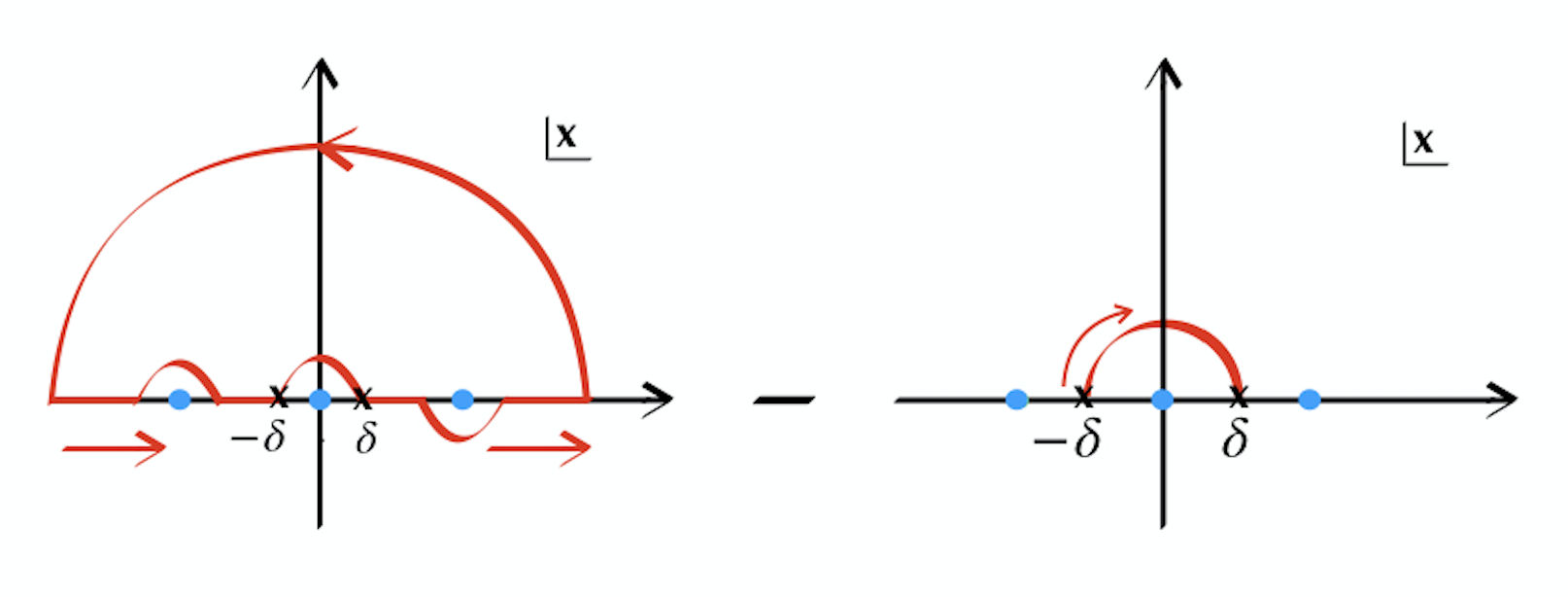}
  \caption{A contour for computing the second term in~Eq.~\eqref{MasslessSpectralGreensFunction1}.}
  \label{fig1}
\end{figure}
\vskip 0.cm
\noindent
The integrand on the right hand side of this expression has 3 poles, those at $x = \pm (pL + i\epsilon)$ and at $x=0$.~The last of these poles is avoided by the $\delta$-prescription in the integral \eqref{MasslessSpectralGreensFunction1} (that is, by the fact that we are integrating up to $-\delta$, and from $+\delta$).~The way we traverse the former two poles is determined by the standard Feynman $i\epsilon$ prescription.~Closing the integration contour by the large semicircle at infinity, as well as the small semicircle between $-\delta$ and $+\delta$, the integral can be readily computed.~The integrand decays sufficiently fast at infinity so that the large semicircle contributes nothing.~The small semicircle, on the other hand, does give a non-zero contribution, which has to be substracted to get the desired term in \eqref{MasslessSpectralGreensFunction1}.~This situation is schematically depicted in Fig.~\ref{fig1}, where we chose to close the contour in the upper half of the complex plane --- closing it in the lower half would lead to the same result, as can be straightforwardly checked.~The contour on the left picks up the single pole at $x = pL + i\epsilon$.~Computing the contribution of this pole and substracting from it the contribution of the semicircle on the right of Fig.~\ref{fig1}, we finally get:
\begin{align}
\Delta_{\nu = 2} (p^2) &=\frac{1}{p^2 + i \epsilon} +  \frac{L^2}{2} \frac{H_2^{(1)} (p L)}{(p L) H^{(1)}_{1} (p L)} - \frac{1}{p^2} \nonumber \\ 
&=  \text{P.V.} \(\frac{L^2}{2} \frac{H_2^{(1)} (p L)}{(p L) H^{(1)}_{1} (p L)}\) - i \pi  \delta(p^2)\,, \label{finalexp}
\end{align}
where the second and the third terms in the first line stem from the left and the right contours in Fig.~\ref{fig1} respectively\footnote{It is straightforward to see that the semicircle on the right panel of Fig.~\ref{fig1} provides a finite contribution to the integral.~To that end, note that $H_2^{(1)}/H_1^{(1)}\propto x^{-1}$ in the limit $x\to 0$, and the integral over the small semicircle $s_1$ is proportional to $\int_{s_{1}} dx/x = - i\pi$.}, while the first term in the second line emerges from 
the three terms in the first line.~Eq.~\eqref{finalexp} gives the final form of the $m_5=0$ amplitude, which exactly agrees with the $m_5\to 0$ limit \eqref{fullamp} of its counterpart in the theory with non-zero $m_5$.~Since the considered scalar has no self interactions, and its interactions with the 5D graviton fluctuations are neglected, 
the above-established continuity of a two-point amplitude establishes continuity of the theory in the bulk 
scalar mass.


\subsection*{The static potential}
It is instructive to understand continuity of the theory in the limit $m_5\to 0$ on a simple physical observable --  the static Newtonian potential between brane sources.~We will start with looking at the large-distance limit of the amplitude \eqref{propagator}, corresponding to $pL \ll 1$.~Furthermore, we will also assume that the bulk mass is small compared to AdS curvature, $\nu - 2 \ll 1$.~One can then expand $\Delta_\nu(p^2)$ as follows
\begin{align}
\Delta_{\nu} (p^2) &\approx \frac{L^2}{(p L)^2 \(1 - (\nu-2)\)- 2 (\nu-2) - \frac{(p L)^4}{2}\, \text{ln}(p L/2 i)}  \,,
\label{expandedAmplitude}
\end{align}
where the corrections to the denominator are of order $(pL)^6, \,(\nu-2)^2\, (pL)^2$ and $(\nu-2)\,(pL)^4$.~Keeping only terms of order $pL$ and $\nu-2$ in \eqref{expandedAmplitude}, this gives rise to the Yukawa potential 
\begin{align} \label{Yukawa}
V_1 (r) = \int \frac{d^3 \mathbf{p}}{(2 \pi)^3}\, e^{i \vec{\mathbf{p}} \cdot \vec{r}}\, \frac{L^2}{-(\mathbf{p} L)^2 - 2 (\nu-2)} \approx -\frac{e^{-m_0 r}}{4 \pi r}\,.
\end{align}
In general, the correction $V_2$ to this expression depends on the relative magnitude of the two expansion parameters.~In what follows, we will focus on deriving two different limits of this correction.\footnote{It is important to note that the term `correction' for the extra term will not be adequate at distances $r \gg m_0$ at which the Yukawa potential is exponentially suppressed.~In such a regine, the extra term $V_2 \propto r^{-7}$ provides the leading contribution, as stressed in the discussion to follow.}

\subsubsection*{Large distances: $r \gg m^{-1}_5$}
At large distances, momentum transfer is the smallest scale in the problem, in particular $pL \ll \nu - 2$, and expanding the amplitude \eqref{expandedAmplitude} in momentum yields:
\begin{align}
V (r \gg m^{-1}_5) = V_1  +  \int \frac{d^3 \mathbf{p}}{(2 \pi)^3} \,\, e^{i {\mathbf{p}} \cdot \mathbf{r}} \,\, \frac{ L^2\, (\mathbf{p} L)^4\, \text{ln}(\vert \mathbf{p}\vert L/2)}{8 (\nu-2)^2} =V_1 - \frac{45 L^5}{4 \pi (\nu-2)^2 r^7}\,\,. \label{verylargedist}
\end{align}
At distances under consideration, $r \gg m_5^{-1}$, the Yukawa potential $V_1$ is contributed mainly by modes with $m\, \gsim\,  m_5$ and is exponentially suppressed.~The leading contribution to the potential is thus given by the second term in \eqref{verylargedist}, as discussed in Ref.~\cite{Dubovsky:2000am}.

\subsubsection*{Small bulk mass: $m_5 \to 0$}

The previous, large distance limit $r \gg m^{-1}_5$ is obviously not consistent with the massless limit of the bulk scalar.~This limit, not considered in \cite{Dubovsky:2000am}, is our main focus in the present section.~When $m_5$ is sent to zero (that is, when $m_5$ is the smallest scale in the problem, but not strictly zero), both the mass and the width of the 4D resonance tend to zero as well, the latter vanishing faster than the former.~The Yukawa potential of Eq.~\eqref{Yukawa} therefore turns into the Newtonian one:
\be
V_1(r)\vert_{m_5 \to 0} = -\frac{1}{4\pi r} \,. \label{Newton}
\ee
While this visually resembles the potential due to the exchange of a 4D massless scalar, it is important to stress that in the setup under consideration there is no localized massless mode in the problem.~The bulk mass $m_5$ is far smaller than any other scale, but it is still not strictly zero -- the KK spectrum therefore does \emph{not} possess a massless state (zero mode), and the potential \eqref{Newton} is due to a linear combination of KK modes with $m > 0$.~The correction to the Newtonian potential $V_1$ can be found by expanding the $\nu\to 2$ limit of the amplitude \eqref{expandedAmplitude} 
 \begin{align}
V_2 (r) &= \frac{L^2}{2}  \int \frac{d^3 {\textbf p}}{(2 \pi)^3} \, e^{i \mathbf{p} \cdot \mathbf{r}}\,  \text{ln}(\vert \mathbf{p}\vert  L/2) = - \frac{L^2}{8 \pi r^3}\,. \label{correction}
\end{align}
Summing up the two limiting contributions in \eqref{Newton} and \eqref{correction} yields the expression, equivalent to what one would get for the static potential in the case of a \emph{strictly massless} bulk scalar.~As we have remarked above, in the latter case there does exist a zero mode in the four-dimensional spectrum, and it leads to the Newtonian force, equivalent to \eqref{Newton}.~Moreover, in the $m_5 = 0$ theory, the gapless continuum of KK modes works exactly as it does in the case of a small but non-zero $m_5$, providing a correction, equivalent to \eqref{correction}.~We have therefore established continuity between the two theories with strictly vanishing $m_5$ and however small, but non-zero $m_5$: the physical spectra of these theories are somewhat different, but all observables---in particular the gravitational potential between brane sources---are perfectly continuous. 

\section{Massive Scalar with a Zero Mode}

Formally, even for non-vanishing bulk mass $m_5$, the bulk equation \eqref{KKequation} for the scalar's 4D KK modes admits a would-be (normalizable) zero mode solution with the $z$-profile proportional to $(\vert z\vert + L)^{-\nu + 2}$.~As already remarked in Sec.~\ref{massivescalar}, however, this solution is incompatible with the \emph{boundary condition} \eqref{KKboundarycondition}, which effectively removes it from the physical KK spectrum.~This observation suggests that the entire theory may be made compatible with the existence of the zero mode---even for non-zero bulk mass---by modifying the \emph{boundary part} of the original scalar action (a similar mechanism has been considered for the case of a massive bulk vector in \cite{Batell:2005wa}).

In what follows, we will show that modifying the theory by an extra scalar mass term, localized at $z=0$ does the job of re-introducing the zero mode on the brane.~Indeed, consider the following theory
\begin{align}
\label{newaction}
S &= S_\phi + \frac{\nu - 2}{2 L} \int_{z=0} d^4 x\, \phi^2\,,
\end{align}
where $S_\phi$ denotes the action of the ``minimal'' massive theory \eqref{action}.~Notice also, that the new boundary term has a ``tachyonic' sign, which, however, does not lead to inconsistency of the theory -- indeed, we will show that the effects of the ``wrong'' sign boundary mass will be overwhelmed by the effects of the ``correct'' sign bulk mass in $S_{\phi}$.~Physically, the tachyonic mass can be thought of as giving an additional attractive contribution to the effective (`volcano') potential that traps the scalar zero mode on the brane.

With the additional boundary term, the dynamics is governed by the modified equation of motion
\begin{align} 
\Big(-\Box + \partial_z^2 - \frac{3 \text{sgn}(z)}{\vert z \vert + L} \partial_z - \frac{(m_5 L)^2}{(\vert z \vert + L)^2} + \frac{ \nu - 2}{L}\,  \delta(z) \Big) \phi (x,z) = -L j(x) \delta(z)\,,
\label{eq2}
\end{align}
which yields the following boundary condition on the brane
\begin{align} \label{boundaryconditiion2}
\Big( \partial_z + \frac{\nu-2}{L} \Big) \phi \vert_{z = 0} = -\frac{L}{2} j(x)\,.
\end{align}
The KK mode wavefunctions can be found by solving the system \eqref{eq2} and \eqref{boundaryconditiion2} in the absence of sources, which establishes that the 4D spectrum of the theory indeed consists of the zero mode scalar, in addition to the continuum of massive KK states! The bulk profiles for these modes read 
\begin{align}
\chi^{(0)}_\nu (z) &= \sqrt{\nu-1} \Big(\frac{L}{\vert z \vert + L} \Big)^{\nu-2}\,, \nonumber \\
\chi^{(m)}_\nu(z) &= \sqrt{\frac{m L}{2}} \Big(\frac{\vert z \vert + L}{L} \Big)^2 \Big[ \,\frac{ J_{\nu-1} (m L) Y_{\nu} (m (\vert z \vert+L))-Y_{\nu-1} (m L) J_{\nu} (m (\vert z \vert+L))}{Y_{\nu-1} (m L)^2 + J_{\nu-1} (m L)^2}\, \Big]\,,
\end{align}
where, to avoid notational clutter, we have kept the same notation for the KK wavefunctions as in the previous section, although the corresponding functions are of course different in the modified theory at hand.~Plugging the KK decomposition $\phi (x, z) = \phi^{(0)} \chi^{(0)}_\nu (z) + \int_0^{\infty} dm L \phi^{(m)} (x) \chi^{(m)}_\nu(z)$ into the original action \eqref{newaction} and using the standard orthonormality properties of the KK wavefunctions, we arrive at the 4D effective theory of the following form
\begin{align}
S =L  \int d^4 x  \Bigg[& \frac{1}{2} (\partial \phi^{(0)})^2 + \sqrt{\nu-1} \,\phi^{(0)} j
\nonumber \\
&+ \int_0^{\infty} dm L \, \left( \frac{1}{2} (\partial \phi^{(m)})^2 - \frac{m^2}{2} (\phi^{(m)})^2 +\chi^{(m)}_\nu (0)\, \phi^{(m)} j\,. \right) \Bigg]\,,
\end{align}
One can see, that all modes are well-behaved despite the ``wrong'' sign of the brane mass term in~\eqref{newaction}.~Moreover, the canonically normalized zero mode and the massive KK modes couple to brane sources with strength, set by the quantities $\sqrt{(\nu - 1)/L}$ and $\chi^{(m)}(0)/\sqrt{L}$ respectively.~Exchange of these modes between brane sources gives rise to the following amplitude
\begin{align}
\Delta (p^2) &= \frac{\nu-1}{p^2 + i\epsilon} + \int_0^{\infty} dm L\, \frac{\vert \chi^{(m)}(0) \vert^2}{p^2 - m^2+i\epsilon} \nonumber \\
&= \frac{\nu-1}{p^2+i\epsilon}  + \frac{2 L^2}{\pi^2} \int_0^\infty d x\,  \frac{1}{(p L)^2 - x^2 + i\epsilon}\,\, \frac{1}{x\, (Y_{\nu-1}(x)^2 + J_{\nu-1}(x)^2)}\,.
\end{align}
By performing manipulations on the second term, very similar to the ones discussed in the previous section, one can integrate over $x$ thereby arriving at the final, closed-form expression for the amplitude
\begin{align}
\Delta (p^2) =\text{P.V.}\,\( \frac{L^2}{2} \frac{H_{\nu}^{(1)}(pL)}{(p L) H^{(1)}_{\nu-1} (p L)} \) - i\pi \,(\nu-1)\, \delta(p^2)\,.
\end{align}
This expression differs by the order of the Hankel functions involved, as well as by the strength of the massless pole, from its counterpart \eqref{finalexp} in the massless bulk theory, described in the previous section. 

\section{de Sitter brane and Cosmology}
\label{dsbranecosmo}
The above discussion has exclusively concerned the case of a flat brane, which requires tuning the brane tension $\lambda$ against the bulk cosmological constant $\Lambda$.~It is known due to Kaloper \cite{Kaloper:1999sm} and Nihei \cite{Nihei:1999mt},
 that upon detuning these two quantities, one can end up with an inflating (de Sitter) brane, instead of a flat one.
 ~The metric of the corresponding spacetime reads \cite{Kaloper:1999sm, Nihei:1999mt}:
\be
\label{metricinit}
ds^2 = \(\cosh\frac{ y}{L} - q \sinh \frac{ y}{L}\)^2 (-dt^2 + e^{2Ht} d\vec x^2) + dy^2\,,\qquad \(q \equiv \sqrt{1+(HL)^2}\)\,,
\ee
where the brane is located at $y=0$ in the given coordinates, $L$ is the curvature radius of the bulk and $H$ is the (constant) Hubble rate on the $dS_4$ spacetime on the brane worldvolume, determined by $\Lambda$ and $\lambda$ -- the precise relation won't be important for our purposes.
\vskip 0.15cm
\noindent
It will prove convenient to perform a change of coordinates
\be
e^{ y/L} = \frac{q+1}{HL}\, \frac{e^{H( z+z_0)}-1}{e^{H( z+z_0)}+1}\,, \qquad z_0 \equiv H^{-1} \ln \frac{q + HL + 1}{q - HL + 1}\,,
\ee
which removes the coordinate singularity in the line element \eqref{metricinit}, putting is into the following form
\be
\label{metricfin}
ds^2 = \frac{(HL)^2}{\sinh^2 H(z +z_0)}\,\(-dt^2 + e^{2Ht}d\vec x^2 + dz^2\)\,.
\ee
In the new coordinates, the brane is located at $z=0$, and we will assume that the theory is invariant under reflections of this coordinate, $z\to -z$, so \eqref{metricfin} should be viewed as describing the brane's ``positive side'', which we will exclusively work with in the following discussion.~On this side of the brane, the $z$ coordinate ranges from $0$ to $\infty$ as  $y$ ranges from zero to a finite value, determined by the location of the coordinate singularity in \eqref{metricinit}, the Rindler horizon. 
Hence, the $z$ coordinate does not cover the entire  space covered by the $y$ coordinate, but 
only its patch ranging from the origin to the Rindler horizon.

While we could continue the spacetime past the horizon to values of $y > L \,\text{ln} \Big( \frac{q+1}{H L} \Big)$, anything happening beyond this region would not effect the observers on the brane. While a signal sent from the brane to the horizon would take a finite proper time to arrive, it would take an infinite amount of time according to an observer on the brane. Furthermore, the boundary conditions for the KK modes are already completely determined at the horizon by the condition of normalizability so we set aside the question of any extension of this spacetime beyond the horizon as irrelevant for our purposes.

\vskip 0.15cm
\noindent
Consider the Euler-Lagrange equation of motion for a massive, $z$-reflection-even bulk scalar $\phi(x, z) = \sigma(x)\, \chi(z)$ in the background \eqref{metricfin}:
\begin{align}
\label{chieq}
\frac{d^2\chi}{du^2} - 3\,\frac{\cosh u}{\sinh u}\, \frac{d\chi}{du} + \(\frac{m^2}{H^2} - \frac{m_5^2 L^2}{\sinh^2 u}\) &=0\,,
\end{align}
where, with the Kaluza-Klein decomposition in mind, we have defined $\Box_4\sigma = m^2\sigma$, as well as $\bar z\equiv z + z_0$ and $u \equiv H\bar z$.~To further simplify notation, we will also define 
\be
A^2 = \frac{m^2}{H^2}\,, \qquad B^2= m^2_5 L^2\,,
\ee
so that the equation of interest \eqref{chieq} becomes
\begin{align}
\label{chieq1}
\chi'' - 3\, \frac{\cosh(u)}{\sinh(u)} \, \chi' + \( A^2 - \frac{B^2}{\sinh^2(u)}\)\, \chi = 0\,, \qquad \chi'\vert = 0\,.
\end{align}
Here, the last equation---the boundary condition on the brane---can be obtained by integrating the bulk equation across the brane and keeping in mind that $\chi$ is a $z$-reflection-even field (again, the vertical stroke denotes evaluation at $u = H z_0$).~A solution to this equation with given $m^2$ (that is, given value of the parameter $A^2$) describes a localized, normalizable mode if it satisfies
\be
\label{normalizability}
\int_0^\infty dz\, \sqrt{g}\, g^{00}\, \chi^2 = \text{finite}\,. 
\ee
This imposes a second boundary condition, this time at $u\to\infty$, on normalizable modes.~In what follows, we will be looking for precisely such localized modes.

A general solution to \eqref{chieq1} reads
\begin{align}
\label{solution}
\chi(u) &=  \cosh(u)^{\frac{1}{2}\(3 - \sqrt{9 - 4 A^2}\)}\, \(\tanh u\)^{2 - \sqrt{4 + B^2}} \nn \\
&\times \,  \Big[C_2\, \cdot {}_2F_1\(a_1, b_1, c_1, t\) -C_1 \cdot \(\tanh u\)^{2\sqrt{4+B^2}}\cdot  {}_2F_1\(a_2, b_2, c_2, t\)  \Big]  \,,
\end{align}
where $C_1$ and $C_2$ are the two integration constants and the coefficients $a_{1,2}, b_{1,2}$ and $c_{1,2}$, together with the $u$-dependent quantity $t$ have been defined as follows
\begin{align}
a_1 &= \frac{1}{4}\(1 + \sqrt{9-4A^2}-2\sqrt{4+B^2}\),
b_1 = \frac{1}{4} \(3 + \sqrt{9-4A^2}-2\sqrt{4+B^2}\),
c_1 = 1-\sqrt{4+B^2}, \nn \\
a_2 &= \frac{1}{4}\(1 + \sqrt{9-4A^2} + 2\sqrt{4+B^2}\),
b_2 =\frac{1}{4} \(3 + \sqrt{9-4A^2} + 2\sqrt{4+B^2}\),
c_2 = 1 + \sqrt{4+B^2} \,,\nn
\end{align}
and $t = \tanh^2 u$.~In order to fix the integration constants $C_1$ and $C_2$, we will need to study the behavior of the solution \eqref{solution} for both large and small values of $u$ -- something we will turn to next.

\subsubsection*{Large $u$}
For large $u$ (corresponding to $t=\tanh^2 u \to 1^-$), and for $c-a-b < 0$ which is the case for the solution \eqref{solution},\footnote{Note that $c_1 - a_1 - b_1 = c_2 - a_2 - b_2 = -\sqrt{9 - 4A^2}/2$\,.} the relevant limit of the hypergeometric function reads
\be
\lim_{t\to 1^{-}} {}_2F_1(a, b, c, t) = (1 - t)^{c - a - b}\,\, \frac{\Gamma(c)\Gamma(a+b-c)}{\Gamma(a)\Gamma(b)}\,,
\ee
which gives in our case:
\be
\lim_{u\to \infty} {}_2F_1(a_{1,2}, b_{1,2}, c_{1,2}, \tanh^2(u)) = \( \cosh u\)^{\sqrt{9-4A^2}}\, \frac{\Gamma(c_{1,2})\Gamma(\sqrt{9-4A^2}/2)}{\Gamma\(a_{1,2})\Gamma(b_{1,2}\)}\,.
\ee
Plugging this expression into the general solution then yields
\be
\label{largex}
\chi(u\gg 1) \simeq \Gamma(\sqrt{9-4A^2}/2)\, \( \cosh u\)^{\frac{1}{2}\(3 + \sqrt{9 - 4 A^2}\)}\,\(C_2\, \frac{\Gamma(c_1)}{\Gamma(a_1)\Gamma(b_1)} - C_1\, \frac{\Gamma(c_2)}{\Gamma(a_2)\Gamma(b_2)}\)\,.
\ee
This correctly reproduces one of the growing modes at large $u$. To see this, we note that when $u \to \infty$, the two independent solutions have the form $\exp (k_{1, 2}\, x)$, where $k_1$ and $k_2$ are the two solutions of the quadratic equation $k^2 - 3kx + A^2=0$; explicitly, $k_{1, 2} = (3 \pm \sqrt{9-4A^2})/2$ and one can see that  \eqref{largex} reproduces the $k_1$-mode, while the $k_2$-mode corresponds to the other, sub-leading solution. The latter solution describes a localized, normalizable mode (with our proper definition of normalizability, given in eq.~\eqref{normalizability}), while the former corresponds to a non-normalizable mode and should thus be removed. To that end, we need to tune $C_1$ and $C_2$ as follows:
\be
\label{relationCs}
C_2 =  C_1 \,\frac{\Gamma(a_1)}{\Gamma(a_2)}\, \frac{\Gamma(b_1)}{\Gamma(b_2)}\, \frac{\Gamma(c_2)}{\Gamma(c_1)}\,.
\ee
With this tuning at hand, one can go ahead and study the behavior of the solution in the opposite limit -- the one corresponding to  $u\ll 1$.
\subsubsection*{Small $u$}
Let us now try to understand whether one can satisfy the boundary condition in eq.~\eqref{chieq1}
\be
\label{boundarycond}
\chi'\vert \equiv \chi'(\epsilon) = 0\,.
\ee
Note, importantly, that the value of $u$ on the brane is small $$\epsilon \equiv Hz_0 \sim HL \ll 1,$$ and let us first try to understand the solution analytically. For $x\ll 1$ our equation becomes
\be
\chi'' - \frac{3}{u}\, \chi' - \frac{B^2}{u^2}\, \chi = 0\,,
\ee
which is solved by
\be
\chi(u\ll1)\simeq \tilde C_1 \, u^{2 - \sqrt{4+B^2}} + \tilde C_2 \, u^{2 + \sqrt{4+B^2}}\,,
\ee
where $\tilde C_{1,2}$ will be expressed in terms of $C_{1,2}$ for our particular solution of interest -- we will give the explicit expressions for these coefficients below. Importantly, neither of the $\tilde C_{1,2}$ automatically vanish for our solution, and the (derivatives of the) two terms can balance each other  to satisfy the boundary condition \eqref{boundarycond} at $u=\epsilon$. Taylor-expanding the solution \eqref{solution} and using \eqref{relationCs}, we have
\begin{align}
C_1^{-1}\, \chi'(\epsilon) &= (2 - \sqrt{4+B^2})\,\frac{\Gamma(a_1)\Gamma(b_1)\Gamma(c_2)}{\Gamma(a_2)\Gamma(b_2)\Gamma(c_1)} \, \epsilon^{-\sqrt{4+B^2}}\,(\epsilon + \mathcal{O}(\epsilon^3))\nn \\ & -(2 + \sqrt{4 + B^2}) \,\epsilon^{\sqrt{4+B^2}}\(\epsilon + \mathcal{O}(\epsilon^3)\) \,.
\end{align}
(The precise expression for $C_1$---not important for the discussion to come---can be found from orthonormality of the AdS mode functions.)~Examining this expression, one can see that there certainly exist possibilities for it to vanish, one of which can be understood as follows: the two terms balance each other for 
\be
\epsilon^{2\sqrt{4+B^2}} =  \frac{2 - \sqrt{4+B^2}}{2 + \sqrt{4+B^2}}\,\,\frac{\Gamma(a_1)\Gamma(b_1)\Gamma(c_2)}{\Gamma(a_2)\Gamma(b_2)\Gamma(c_1)} \,;
\ee
under our assumptions, $\epsilon$ is small, and so should be the right hand side of this expression. There seems to be at least the following possibility for this: $2 - \sqrt{4+B^2}$ is small if $B$ is small and on top of that $\Gamma(c_1)$ is large, which makes the left hand side small for a generic choice of $A$. Of course, for this to be true, one has to check that there are no surprises at higher orders in Taylor expansion in $\epsilon$.~We can prove the absence of such surprises numerically by examining how the \textit{exact} (unexpanded) $C_1^{-1}\, 
\chi'(x)$ behaves for different $A$ and $B$.
\begin{figure}
  \includegraphics[width=12cm]{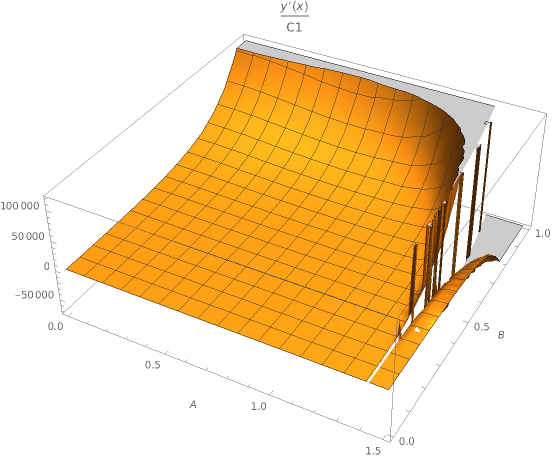}
  \caption{The dependence of $C_1^{-1} y'(x)$ on the parameter $A$ for $B=\epsilon = 10^{-4}$.}
  \label{fig1}
\end{figure}
\subsection*{(Numerical) study of the exact expression for $y'(x)$}
The closed-form expression for $C_1^{-1}\, \chi'(x)$ is quite cumbersome and we will not reproduce it here.~Instead, let us choose a representative value for $\epsilon$, such as $\epsilon=10^{-4}$, and numerically explore the dependence of this expression on $A$ and $B$.
\vskip 0.15cm
\noindent
This dependence is shown, in the form of a 3D plot, on Fig.~1 (where irregularities arise whenever $a_1$ or $b_1$ cross negative integers). One can see that the function of interest comes close to zero for small B, but it is not clear enough whether it actually crosses zero.~To see whether $C_1^{-1}\, \chi'(x)$ crosses zero, let us choose a particular value $B = 10^{-4}$ and see if a value for $A$ exist such that this function vanishes.
\begin{figure} 
\includegraphics[width=12cm]{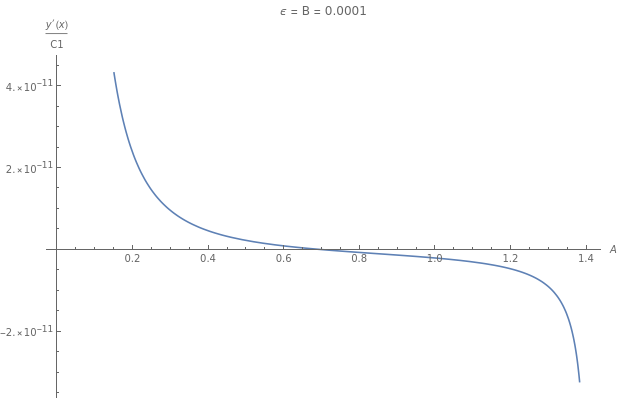}
  \caption{The dependence of $C_1^{-1} y'(x)$ on the parameters $A$ and $B$. }
  \label{fig1}
\end{figure}
\vskip 0.15cm
\noindent
The dependence of $C_1^{-1}\, \chi'(x)$ on $A$ for $\epsilon=B=10^{-4}$ is shown on Fig.~2.
It is clear that a value of $A$ exists, such that the desired function smoothly crosses zero  --  exactly like we predicted from the Taylor expansion argument, given above. (We emphasise again that the present argument is not using any approximation -- we are working with exact expressions here.)
\vskip 0.15cm
\noindent
This establishes the presence of a localized mode for the case of a de Sitter brane. 
It is straightforward to see that  continuum of the KK excitations in this case starts above a gap determined
by the 4D curvature scale:
\beq
m_{KK}^2\ge {9H^2\over 4}\,,
\label{KKgap}
 \eeq
and the localized mode resides within the  gap. It is difficult to  calculate the mass of the localized
mode. However, one can deduce scaling dependence of the mass of this mode on other parameters
by approximately matching asymptotic solutions. Doing so, one obtains for the mass of the 
localized mode, $m_* \sim m/ (HL)$, in the approximation when $HL<<1$ and  $m_*<< 3H/2$.
Thus, in the above regime of the parameter space the mass of the localized mode is significantly 
larger than the mass of the bulk mode, $m$, but is smaller than the curvature scale $H$.

\subsection*{Comments on Cosmology}

We will outline here how the 5D massive scalar in $AdS_5/Z_2$  can be 
used to describe an accelerated expansion of the universe. The rate of the 
acceleration will depend on the scalar mass, and  therefore, choosing the value 
of the mass one can either make the expansion suitable  for  inflation, or for the 
late time acceleration.

To begin with consider the 5D massive scalar embedded in the geometry 
with a tuned brane and bulk, as in Section 2,  so that the 4D brane world-volume  
spacetime is flat.~It remains flat as long as the scalar is in its vacuum state,  $\phi=0$. 

Let us now imagine that at some  earlier time  the initial value of of the scalar was nonzero, 
$\phi=\phi_0$, with its time derivative being negligible. At that initial moment the 
scalar would add to the bulk  energy density a positive quantity 
\beq
\Delta E = {1\over 2} m_5^2 \,\phi_0^2\,.
\label{DeltaE}
\eeq
For simplicity,  we will assume that this energy density is  
less in its magnitude than the bulk  AdS negative energy density; 
thus,  the  quadratic scalar potential will  lead to a  reduction of the magnitude of 
the negative energy density in the bulk. As a result, the bulk  energy density and 
the brane tension will  no longer be tuned, and  the brane world-volume will  at that time 
moment acquire a positive space-time curvature \cite {Kaloper:1999sm, Nihei:1999mt}
\beq
H = \sqrt {{\Delta E \over M_5^3}}\,,
\label{H}
\eeq
where $M_5$ denotes the (specifically normalized) Planck mass of the 5D theory. 
Let us furthermore assume that  after the initial moment the 5D scalar field is going to 
roll down its quadratic potential slowly  during some reasonable classical 
interval of time.  As a result of the slow, roll the curvature of 
the 5D spacetime will change slowly too, and so will  the curvature of the 4D (quasi) 
de Sitter universe. This expansion can potentially describe either inflation in the early universe, or
the dark energy driven acceleration, depending on the values of the scalar mass and $\phi_0$.
 
 While we will not pursue the detailed studies of these cosmologies here,
 we point out  two peculiarities of the proposed scheme: first the 4D Planck 
 constant will also be changing in time as the 5D scalar rolls down its slope. 
 Second, the 4D  fluctuations, as shown  in Section~\ref{dsbranecosmo}, will consist 
 of one localized  massive mode below the gap and and the  KK 
 continuum  above a gap.  If applied to inflation, The localized mode  will 
 then be responsible for density perturbations; its mass is parametrically different 
 from the mass of the 5D field that's rolling down. Thus, it might  be interesting to 
 work out  the details of such a cosmological model.

\subsection*{Acknowledgments}

We thanks Nemanja Kaloper for useful correspondence. GG is supported by  the NSF grant PHY-1915219.~Work of DP  and DO at NYU was supported by the Simons Foundation under the "Origins of The Universe" program.

\bibliographystyle{utphys}
\bibliography{scalar} 
\end{document}